\documentclass[11pt]{article}

\usepackage[utf8]{inputenc}
\usepackage[T1]{fontenc}
\usepackage[a4paper,margin=1in]{geometry}
\usepackage{newtxtext}
\usepackage{newtxmath}
\usepackage{graphicx}
\usepackage{booktabs}
\usepackage{array}
\usepackage{tabularx}
\usepackage{longtable}
\usepackage{ltablex}
\usepackage{multirow}
\usepackage{makecell}
\usepackage{adjustbox}
\usepackage{amsmath}
\usepackage[font=small,labelfont=bf,labelsep=period]{caption}
\usepackage{subcaption}
\usepackage{float}
\usepackage{placeins}
\usepackage{needspace}
\usepackage{enumitem}
\usepackage[table]{xcolor}
\usepackage{hyperref}
\usepackage{cite}
\usepackage{algorithm}
\usepackage{algpseudocode}

\DeclareUnicodeCharacter{2010}{-}
\DeclareUnicodeCharacter{2011}{-}
\DeclareUnicodeCharacter{2012}{--}
\DeclareUnicodeCharacter{2013}{--}
\DeclareUnicodeCharacter{2014}{--}
\DeclareUnicodeCharacter{2212}{-}
\DeclareUnicodeCharacter{00B5}{$\mu$}
\DeclareUnicodeCharacter{00FC}{\"u}
\DeclareUnicodeCharacter{00E9}{\'e}
\DeclareUnicodeCharacter{00E1}{\'a}
\DeclareUnicodeCharacter{00ED}{\'i}
\DeclareUnicodeCharacter{00F3}{\'o}

\newcolumntype{Y}{>{\centering\arraybackslash}X}
\newcolumntype{L}[1]{>{\raggedright\arraybackslash}p{#1}}
\newcolumntype{C}[1]{>{\centering\arraybackslash}p{#1}}
\definecolor{groupgray}{gray}{0.92}
\definecolor{rowgray}{gray}{0.95}
\hypersetup{
  colorlinks=true,
  citecolor=green!45!black,
  linkcolor=red!60!black,
  urlcolor=blue!60!black
}

\begin{document}

\title{ChargeBD: Character-Aware Heterogeneous Agent Reasoning for Guided Engineering in Battery Development}

\author{
  Rui Huang\textsuperscript{1,\dag},\quad
  Zekun Jiang\textsuperscript{1,\dag},\quad
  Mengran Hou\textsuperscript{1},\quad
  Xingyu Niu\textsuperscript{1},\quad
  Yuqiang Li\textsuperscript{2},\\[0.3em]
  Qinying Gu\textsuperscript{2,*},\quad
  Tianhang Zhou\textsuperscript{1,3,*}
  \\[0.8em]
  \small\textsuperscript{1}College of Energy Innovation, State Key Laboratory of Heavy Oil Processing,\\
  \small China University of Petroleum (Beijing), Beijing 102249, China\\[0.3em]
  \small\textsuperscript{2}Shanghai Artificial Intelligence Laboratory, Shanghai, China\\[0.3em]
  \small\textsuperscript{3}ZH Energy Storage Technology (Beijing) Co., Ltd., Beijing, China\\[0.6em]
  \small\textsuperscript{\dag}These authors contributed equally.\quad
  \small\textsuperscript{*}Corresponding authors.\\
  \small\texttt{guqinying@pjlab.org.cn},\quad\texttt{zhouth@cup.edu.cn}
}
\date{}
\maketitle

\section*{Abstract}

Redox-flow battery (RFB) research poses constrained, multi-scale reasoning tasks across molecular design, materials, components, operation, system management and safety. We introduce ChargeBD, a persona-conditioned heterogeneous-agent framework for energy-storage R\&D. Starting from a 50-question RFB-specific task set, we constructed the 500-question ESS-LLM Benchmark, selected DeepSeek-V3-Plus as a shared base model, and evaluated 16 MBTI-inspired persona agents as prompt-defined reasoning patterns rather than psychometric instruments. ChargeBD combines task-persona matching, dual-path on-demand activation, parallel generation, cross-review, disagreement convergence and multi-level fusion. On a 100-task core validation set, dynamic activation achieved a higher overall score than a fixed quartet while reducing token use by 48.2\% and inference time by 50.2\%, with 2.6 activated agents on average. The ChargeBD Web Platform exposes task decomposition, persona activation, concurrent reasoning and synthesis as a transparent, auditable workflow. These results support resource-aware computational reasoning workflows for energy-storage R\&D, not experimental validation of battery performance.

\textbf{Keywords:} redox flow batteries; energy storage; large language models; multi-agent systems; persona-conditioned reasoning; heterogeneous agents; battery development

\section{Introduction}

Large language models (LLMs) are increasingly used as scientific reasoning interfaces rather than only text generators. In chemistry, materials science, and battery research, LLMs have been used for literature mining, structured data extraction, domain knowledge organization, and battery-oriented reasoning support \cite{ref01,ref02,ref03,ref04,ref05}. Molecular and materials design studies further show that language models and generative AI workflows can assist candidate generation and design-space exploration \cite{ref06,ref07}. Broader AI-for-science and autonomous-chemistry examples further contextualize these workflows \cite{ref08,ref09}. Tool-augmented and multi-agent studies extend this role toward task planning, tool orchestration, intermediate review, and coordinated scientific workflows \cite{ref10,ref11,ref12,ref13,ref14,ref15,ref16}. These applications show that LLMs can organize distributed knowledge and translate research intent into reviewable workflows. However, they also expose a limitation: a fluent response is not necessarily a constraint-consistent engineering decision, especially when the task requires multi-step reasoning, domain-specific verification, and adaptation across heterogeneous scientific objectives \cite{ref17,ref18,ref19}.

This limitation is consequential in long-duration, grid-scale energy storage, where value depends on more than cell-level energy density or round-trip efficiency. Storage technologies must also be evaluated by duration, safety, lifetime, siting flexibility, capital cost, controllability, and reliability under uncertain renewable output \cite{ref20}. Redox flow batteries (RFBs) are a representative stationary-storage platform because the electrochemical power unit and externally stored electrolyte inventory can be designed with partial independence, enabling flexible scaling of power and energy capacity \cite{ref21}. Their promise is tied to safety, long cycle life, modular scalability, and low-cost deployment pathways rather than to a single material descriptor \cite{ref21,ref22}. This makes RFBs suitable for evaluating AI-assisted reasoning under industrially meaningful constraints.

RFB research is not a single-material optimization problem. Practical development spans redox-active molecular design, electrolyte formulation, electrode kinetics, membrane selectivity, stack and flow-field architecture, flow management, state estimation, control, reliability, and safety protection. RFB reviews and modeling studies emphasize that these layers are coupled across molecular, cell, stack, and plant scales \cite{ref21,ref23}. Component-level studies further illustrate how electrolyte composition, electrode kinetics, membranes, flow fields, and system energy losses shape feasible designs \cite{ref24,ref25,ref26,ref27,ref28}. Thus, an apparently plausible design may fail because it violates transport constraints, manufacturability, operating safety, or system-level energy-loss boundaries. Additional background on RFB multiscale coupling is provided in Supplementary Figure~S19.

\begin{figure}[!htbp]
\centering
\includegraphics[width=0.98\textwidth,height=0.64\textheight,keepaspectratio]{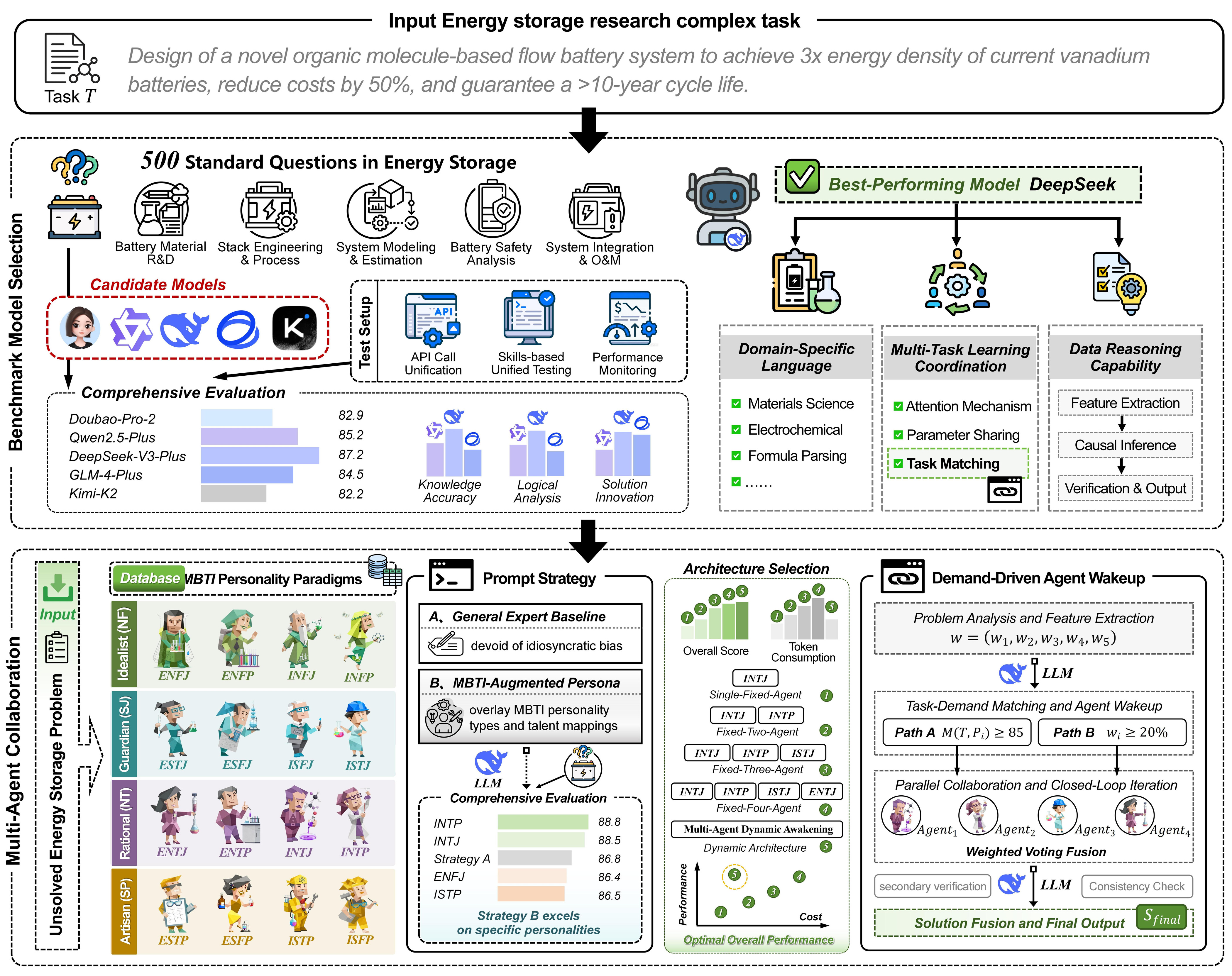}
\caption{\textbf{ChargeBD framework for character-aware heterogeneous-agent reasoning in battery development.} The figure presents the overall ChargeBD workflow, including task parsing, MBTI-inspired persona conditioning, multi-agent collaboration, and fused decision support for constrained energy-storage research tasks.}
\label{fig:personamas_framework}
\end{figure}

The same full-chain structure creates differentiated cognitive demands. Molecular and electrolyte design often require exploratory hypothesis generation, whereas electrode, membrane, flow-field, and stack questions require mechanism tracking, parameter feasibility, and awareness of coupled transport and electrochemical processes \cite{ref23,ref24,ref25,ref26,ref27,ref28}. System-control and state-estimation tasks require consistency with dynamic models, while safety and scale-up questions require conservative reasoning about failure modes, margins, and operational risk \cite{ref23,ref28,ref29,ref30}. RFB research therefore requires the ability to shift reasoning style across exploratory, mechanistic, quantitative, risk-aware, and integration-oriented tasks.

AI and machine-learning methods have become increasingly visible in scientific and engineering workflows, including molecular discovery, materials screening, and model-based optimization in electrochemical energy-storage research \cite{ref06,ref29,ref30,ref31,ref32}. Recent LLM-related studies further suggest that language models can help organize literature-derived knowledge, chemical information, battery-domain evidence, and task-level reasoning steps in scientific workflows \cite{ref01,ref02,ref03,ref04,ref05}. Tool-augmented chemistry systems additionally show how language models can be connected to domain operations and external modules \cite{ref11,ref33}. These studies indicate that LLMs can serve as useful knowledge-processing and reasoning-support tools for scientific research. However, most existing applications still emphasize information extraction, knowledge organization, tool use, or task-specific assistance, leaving the question of how an LLM-based system should adjust its reasoning strategy less explored when the target problem changes across a complex engineering workflow.

This issue becomes especially relevant in RFB research, where useful answers must remain compatible with coupled electrochemical, materials, transport, safety, and system-level constraints \cite{ref21,ref23,ref29,ref30}. Component-level constraints enter through electrolyte formulation, electrode and membrane behavior, flow-field architecture, and system energy-loss boundaries \cite{ref24,ref25,ref26,ref27,ref28}. These constraints are not encountered uniformly across the research chain; they appear with different priorities in molecular design, component optimization, stack operation, system modeling, and safety assessment. As a result, the required reasoning style also changes with task type. For example, a molecular-design task may emphasize redox potential, solubility, and stability, whereas modeling, stack-operation, or safety tasks require mechanistic consistency, parameter awareness, conservative risk assessment, or multi-objective trade-off reasoning. Therefore, the key difficulty is not only whether an LLM can provide relevant knowledge, but whether its reasoning process can adapt to different cognitive demands within the RFB research chain.

Existing agentic and multi-agent studies provide useful precedents for structuring LLM-based scientific workflows. Tool-augmented chemistry agents show that external tools and expert-designed modules can compensate for some limitations of LLMs in chemistry-specific operations \cite{ref11,ref16,ref33}. Scientific multi-agent frameworks further show that role division, intermediate review, and workflow-level coordination can support more organized hypothesis generation and evaluation \cite{ref10,ref12,ref13,ref14,ref15}. These studies motivate structured collaboration, but the RFB setting raises a more task-specific question: how should a system decide which reasoning roles are needed when the problem shifts among exploratory design, mechanistic analysis, engineering verification, and safety-oriented decision making?

Persona prompting offers one way to introduce controllable reasoning-style differences within the same base model. Recent work on role-playing agents, personalization, and persona-conditioned LLMs indicates that personas can guide model behavior according to specific contexts, roles, and task expectations \cite{ref34,ref35,ref36,ref37}. In this study, MBTI is used as a structured typological vocabulary for organizing prompt-induced decision styles, drawing only on its type vocabulary and four-dimensional preference structure \cite{ref38,ref39}. Its prior use in engineering-team learning contexts further suggests that, when used descriptively, such type descriptors can serve as tools for organizing collaboration preferences and decision heuristics \cite{ref40}. MBTI-inspired persona agents are designed as role templates with different reasoning emphases, such as exploratory abstraction, theoretical modeling, structured verification, system-level organization, and risk-aware analysis. Following this task-demand perspective, ChargeBD connects task-demand dimensions with persona-capability dimensions and activates suitable persona-conditioned agents on demand rather than relying on a single fixed prompt or a fixed collaboration group.

\begin{figure}[!htbp]
\centering
\includegraphics[width=0.86\textwidth,height=0.52\textheight,keepaspectratio]{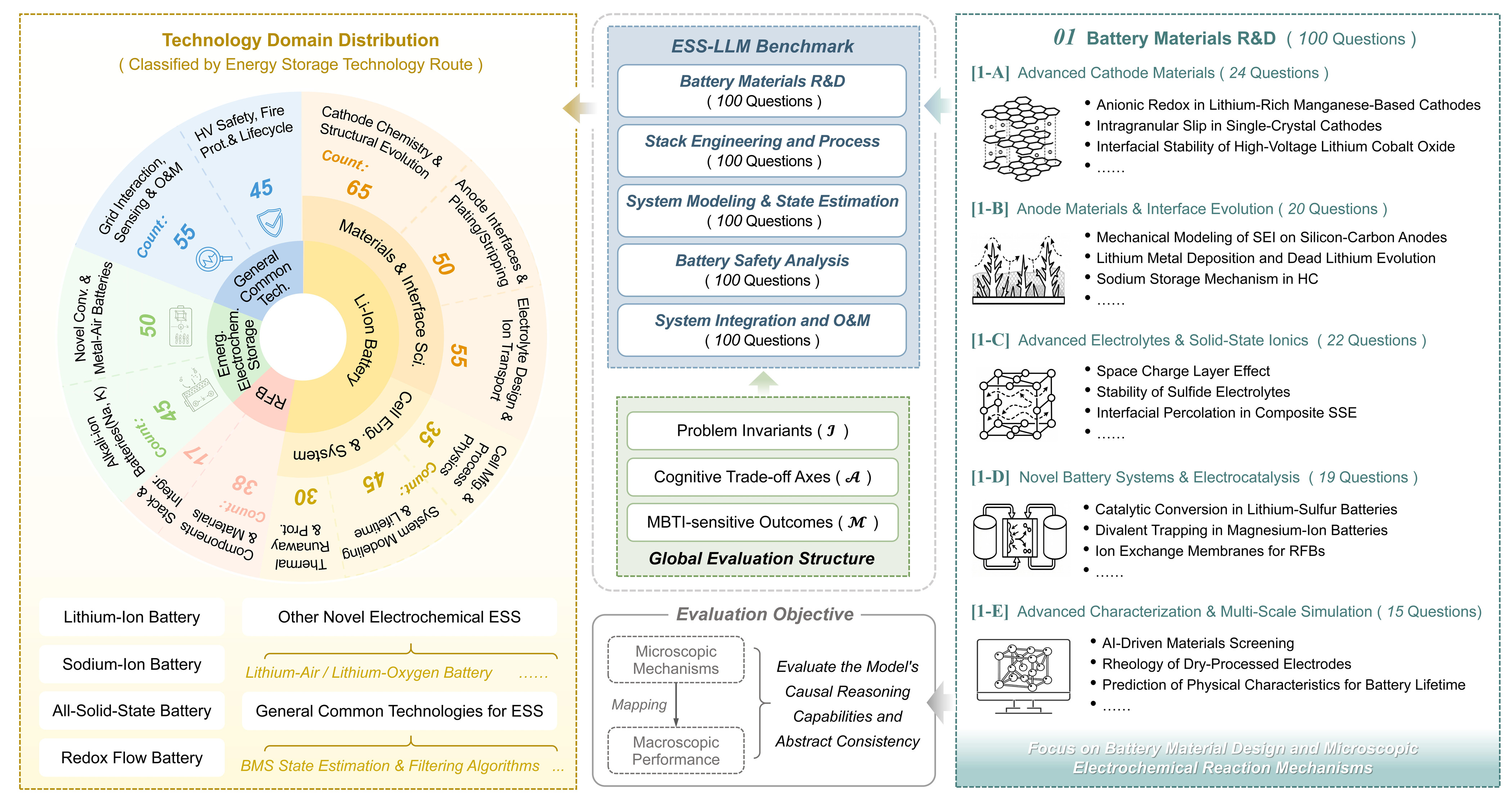}
\caption{\textbf{Hierarchical topic organization and category distribution of the ESS-LLM Benchmark.} The figure shows the construction of the ESS-LLM Benchmark from major energy-storage task categories to subtopics and representative problem instances, supporting balanced evaluation across materials R\&D, stack engineering, system modeling, safety analysis, and system integration.}
\label{fig:ess_llm_benchmark}
\end{figure}

This framing shifts the research question away from identifying a single best persona and toward determining which reasoning capabilities each task requires. A set of 16 persona prompts alone is insufficient if all agents are always activated or if fixed groups are used regardless of task demand. ChargeBD therefore links five task-demand dimensions to six persona-capability dimensions and supports on-demand activation through matching-score and dimension-driven paths. The ESS-LLM Benchmark provides the task substrate by organizing RFB and broader energy-storage problems across materials R\&D, stack engineering, system modeling, safety analysis, and integration-oriented reasoning, as shown in Figure~\ref{fig:ess_llm_benchmark}.

Building on this motivation, ChargeBD integrates 16 MBTI-inspired persona agents, a persona-capability matrix, task-persona matching, dual-path activation, parallel reasoning, cross-review, dispute convergence, and multi-level answer fusion into a task-adaptive workflow. The ChargeBD framework is summarized in Figure~\ref{fig:personamas_framework}, the on-demand activation mechanism is shown in Figure~\ref{fig:wakeup_mechanism}, and the detailed evaluation-control and prompt-strategy workflow is provided in Supplementary Figure~S20. The accompanying ChargeBD Web Platform serves as the implementation-level outcome of this framework, translating ChargeBD into a configurable, transparent, and auditable interactive interface for reasoning tasks in energy-storage engineering research. It is not treated as the source of benchmark scores or as experimental validation of battery performance.

Accordingly, this study makes four contributions. First, it frames RFB research as a constrained, multi-scale, multi-objective energy-storage R\&D setting and constructs the ESS-LLM Benchmark for evaluating reasoning under engineering boundaries. Second, it defines MBTI-inspired persona agents as structured cognitive-bias templates on a shared base model, enabling controlled comparison of prompt-induced reasoning styles without psychometric claims. Third, it proposes task-persona matching and dual-path activation connecting five task-demand dimensions with six persona-capability dimensions. Fourth, it evaluates an on-demand persona-conditioned multi-agent workflow and presents the ChargeBD Web Platform as a transparent, configurable interface for AI-assisted energy-storage research.

\FloatBarrier

\section{Methods}

\subsection{RFB task set and ESS-LLM Benchmark}

This study uses RFB R\&D as the starting point for domain-specific benchmark construction because it spans molecular design, electrolyte regulation, electrode and membrane materials, stack operation, system management, and safety analysis. We first constructed a 50-question RFB-specific task set covering five categories, with each task annotated for objectives, physicochemical constraints, and evaluation priorities. The full RFB-specific task classification is provided in Supplementary Table~S1.

\begin{figure}[!htbp]
\centering
\includegraphics[width=0.98\textwidth,height=0.64\textheight,keepaspectratio]{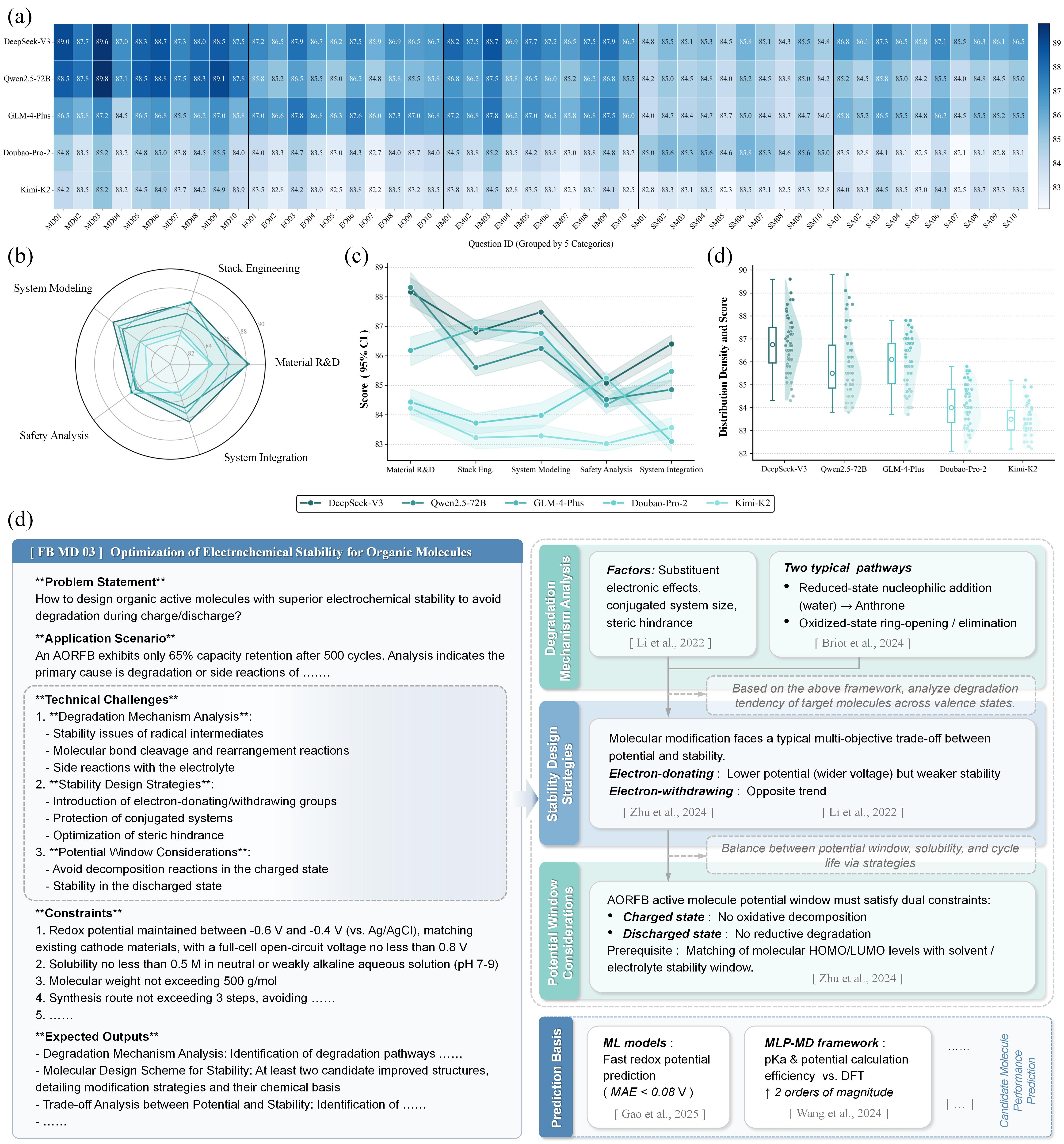}
\caption{\textbf{RFB task-set construction and baseline model selection.} The merged figure presents the constrained RFB task-set design and the RFB-specific comparison used to select DeepSeek-V3-Plus as the shared baseline model.}
\label{fig:rfb_molecular_task}
\end{figure}

The 50-task set embeds realistic constraints into each problem. For example, an organic RFB molecular-design task requires candidate redox-active molecules to balance redox potential, solubility, stability, synthetic feasibility, cost, and environmental safety. The representative input in Figure~\ref{fig:rfb_molecular_task} illustrates how application scenarios, degradation mechanisms, physicochemical boundaries, and multi-objective requirements are combined so that the benchmark evaluates constrained design reasoning rather than only factual recall.

In these tasks, each prompt usually contains an application scenario, a failure or degradation phenomenon, target indicators, and limiting conditions. Model outputs are first constrained by problem invariants, $\mathcal{I}$: answers must not violate thermodynamic laws, electrochemical boundaries, reaction feasibility, safety red lines, or explicit engineering specifications. An output that violates these invariants is treated as factually incorrect even if its language is fluent.

The RFB-specific set is then expanded into the 500-question ESS-LLM Benchmark by generalizing the same constraint-oriented task structure to materials R\&D, stack engineering, system modeling, safety analysis, and system integration. Each major category contains 100 questions and is divided into subtopics to support balanced persona-agent evaluation. The full taxonomy and task objectives are provided in Supplementary Table~S2.

Tasks in the ESS-LLM Benchmark are not only classified by technical topic but also represented by task-demand weights. To support task-persona matching, each task is represented as a five-dimensional task-demand weight vector:

\begin{equation}
\mathbf{w}_{\mathrm{prob}}=(w_1,w_2,w_3,w_4,w_5),\sum_{i=1}^{5}w_i=1
\label{eq:task_vector}
\end{equation}

Here, $w_i$ denotes the demand weight on the $i$-th task dimension. This representation allows one problem to contain materials design, engineering implementation, model derivation, safety boundaries, and system integration simultaneously. For example, a high-energy-density RFB molecular-design task is mainly materials-oriented, but synthetic feasibility, electrolyte compatibility, long-term stability, and system safety introduce nonzero weights in other dimensions. The five-dimensional task-weight representation used in ESS-LLM is defined in the surrounding text, and feature-annotation statistics are provided in Supplementary Table~S3.

\subsection{Persona-agent design and task-persona matching}

In ChargeBD, persona-conditioned reasoning is implemented through MBTI-inspired persona prompts, which are used as structured cognitive-bias templates rather than psychometric representations of real personalities. MBTI types are therefore used as computational heuristics that condition the same base model with different reasoning preferences, risk orientations, and decision priorities.

Two prompt strategies distinguish generic professional prompting from persona-conditioned prompting. Strategy A sets the base model as a senior all-around energy-storage expert and emphasizes scientific rigor, logical consistency, and multi-scale reasoning without a persona bias. Strategy B adds a specific MBTI type, cognitive-feature description, and value sequence on the same professional base. The prompt-template comparison in Supplementary Figure~S20 defines this controlled A/B contrast, and the full prompts are provided in Supplementary Note~S1.

Based on Strategy B, this study constructs 16 MBTI-inspired persona agents. Each agent shares the same base model, task input, and evaluation protocol, differing only in the system-level persona prompt. Performance differences are therefore interpreted as prompt-conditioned reasoning-pattern differences under a unified model capability.

To quantify persona-task adaptation, this study maps the 16 persona agents into a six-dimensional persona-capability space. The persona-capability matrix is denoted as:

\begin{equation}
\mathbf{C}_{\mathrm{cap}}=
\begin{bmatrix}
c_{1,1} & c_{1,2} & \cdots & c_{1,6}\\
c_{2,1} & c_{2,2} & \cdots & c_{2,6}\\
\vdots & \vdots & \ddots & \vdots\\
c_{16,1} & c_{16,2} & \cdots & c_{16,6}
\end{bmatrix},\mathbf{C}_{\mathrm{cap}}\in\mathbb{R}^{16\times6}
\label{eq:capability_matrix}
\end{equation}

Here, $c_{i,j}$ denotes the score of the $i$-th persona agent on the $j$-th capability dimension: innovation insight, material-design ability, theoretical modeling, logical completeness, professional accuracy, and safety-analysis ability. The complete $16\times6$ matrix is reported in Table~\ref{tab:persona_matrix} and supplies capability-side inputs for matching score computation.

We define the capability-demand vector as

\[
\mathbf{w}'=(w'_1,w'_2,\ldots,w'_6).
\]

The five-dimensional task-demand weights and six-dimensional persona-capability dimensions are not one-to-one. Materials-R\&D tasks require both innovation insight and material-design ability, while safety-analysis tasks require safety-analysis ability, logical completeness, and professional accuracy. We therefore use a linear mapping to convert \(\mathbf{w}_{\mathrm{prob}}=(w_1,w_2,w_3,w_4,w_5)\) into the six-dimensional capability-demand vector \(\mathbf{w}'=(w'_1,w'_2,\ldots,w'_6)\), as defined in Table~\ref{tab:task_capability_mapping}.

\Needspace{0.28\textheight}
\begin{table}[!htbp]
\centering
\setcounter{table}{0}
\caption{\textbf{Mapping from task dimensions to capability demands.}}
\label{tab:task_capability_mapping}
\small
\renewcommand{\arraystretch}{1.12}
\begin{tabularx}{\textwidth}{L{0.20\textwidth} C{0.29\textwidth} L{0.43\textwidth}}
\toprule
\textbf{Dimension} & \textbf{Mapping formula} & \textbf{Main contributing task dimensions} \\
\midrule
Innovation insight & \(w'_1=0.6w_1+0.4w_3\) & Materials R\&D (60\%), system modeling (40\%) \\
Material-design & \(w'_2=0.8w_1+0.2w_2\) & Materials R\&D (80\%), stack engineering (20\%) \\
Theoretical modeling & \(w'_3=0.6w_3+0.3w_2+0.1w_5\) & System modeling (60\%), stack engineering (30\%), system integration (10\%) \\
Logical completeness & \(w'_4=0.5w_4+0.5w_5\) & Safety analysis (50\%), system integration (50\%) \\
Professional accuracy & \(w'_5=0.5w_2+0.5w_4\) & Stack engineering (50\%), safety analysis (50\%) \\
Safety-analysis ability & \(w'_6=0.9w_4+0.1w_1\) & Safety analysis (90\%), materials R\&D (10\%) \\
\bottomrule
\end{tabularx}
\end{table}

After obtaining $\mathbf{w}'$, the demand-capability matching score between task $T$ and persona agent $P_i$ is defined as:

\begin{equation}
M(T,P_i)=\sum_{j=1}^{6}w'_j c_{i,j}
\label{eq:matching_score}
\end{equation}

Here, $w'_j$ is the weight of the $j$-th capability demand, and $c_{i,j}$ is the score of persona agent $i$ on that dimension. The matching score is a weighted sum rather than an embedding or distance measure; a higher value indicates stronger fit between persona capability and task demand.

To decide whether a persona should participate in subsequent reasoning, we use a score-driven threshold of \(\theta_{\mathrm{score}}=100\), selected from the statistical relationship between matching-score intervals and actual task scores; full threshold statistics are provided in Supplementary Note~S2. This matching interface then drives on-demand activation.

\subsection{On-demand activation and fusion mechanism}

After task-persona matching is defined, ChargeBD uses a dual-path on-demand activation mechanism to decide which persona agents should participate in each task. Unlike fixed configurations, which call the same agents for all tasks, the mechanism runs an overall matching-score path and a dimension-weight path in parallel. Figure~\ref{fig:personamas_framework} situates this activation step within the full ChargeBD workflow, linking task-persona matching to on-demand activation, cross-review, and multi-level fusion.

\begin{figure}[!htbp]
\centering
\includegraphics[width=0.96\textwidth,height=0.82\textheight,keepaspectratio]{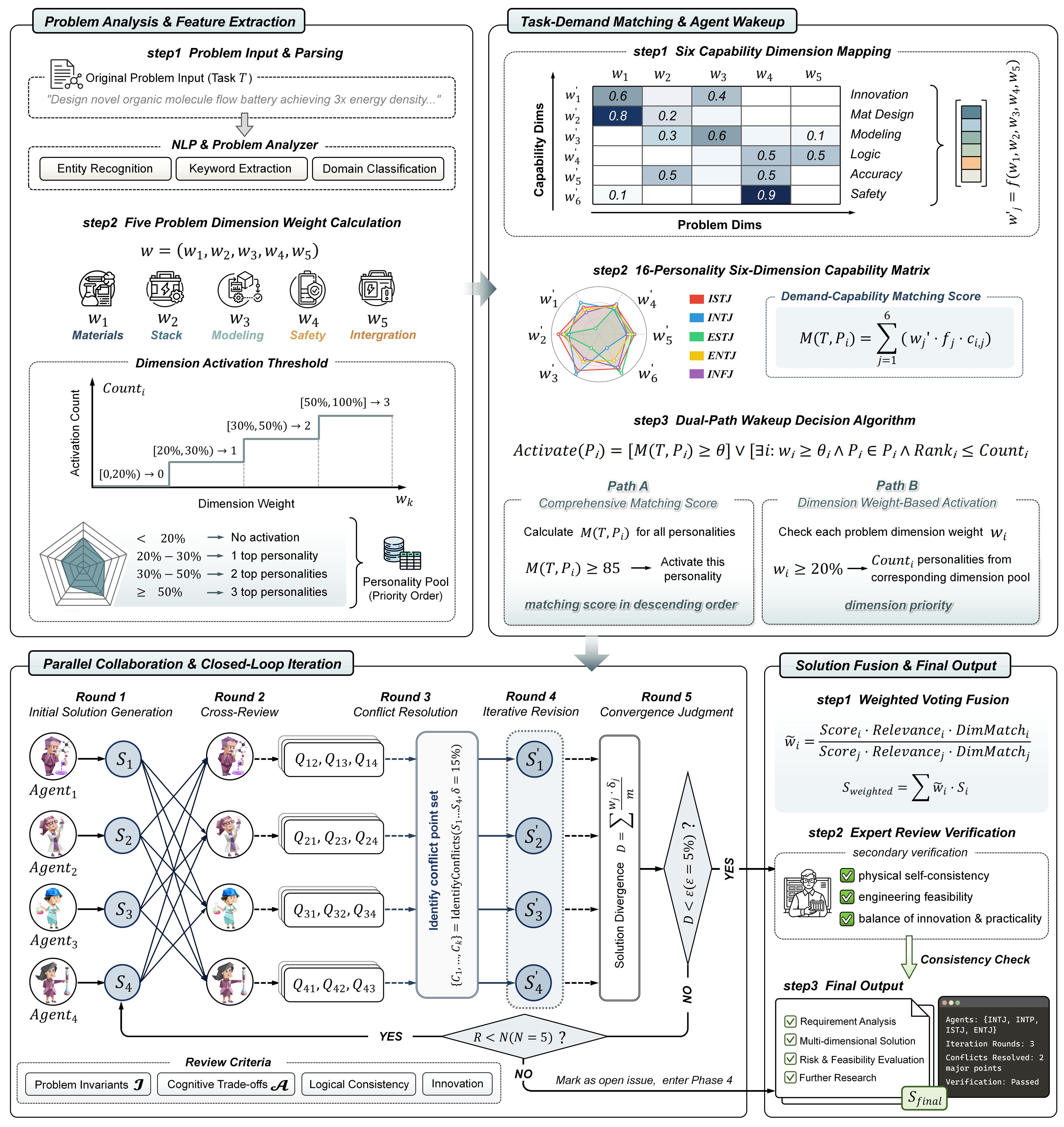}
\caption{\textbf{Demand-driven multi-agent on-demand activation mechanism and dynamic architecture selection.} The figure describes the demand-driven on-demand activation mechanism used by ChargeBD, including task-persona matching, conversion from task-demand weights to capability-demand weights, score-driven activation, dimension-driven activation, and dynamic architecture selection under the maximum activation constraint.}
\label{fig:wakeup_mechanism}
\end{figure}

The demand-driven on-demand activation mechanism in Figure~\ref{fig:wakeup_mechanism} specifies how the two activation paths are converted into a dynamic architecture selection step after task-persona matching.

The first path is score-driven activation, which selects persona agents whose overall matching score exceeds the threshold:

\begin{equation}
\mathrm{Activate}_A(P_i)=
I\!\left[M(T,P_i)\geq\theta_{\mathrm{score}}\right]
\label{eq:activate_a}
\end{equation}

where \(I[\cdot]\) denotes the indicator function and \(\theta_{\mathrm{score}}=100\) is the matching-score activation threshold.

The second path is dimension-driven activation, which selects the highest-priority agents in a dimension-specific persona pool when a task dimension has a high weight:

\begin{equation}
\begin{aligned}
\mathrm{Activate}_B(P_i)
&=I\!\left[
\exists k:\,
w_k\geq\theta_{\mathrm{dim}},\,
P_i\in\mathrm{Pool}_k,
\mathrm{Rank}_k(P_i)\leq\mathrm{Count}(w_k)
\right]
\end{aligned}
\label{eq:activate_b}
\end{equation}

Here, $\theta_{\mathrm{dim}}=20\%$ is the minimum dimension weight required to trigger activation, $\mathrm{Pool}_k$ is the persona pool adapted to the $k$-th task dimension, $\mathrm{Rank}_k(P_i)$ is the priority rank of persona agent $P_i$ in that pool, and $\mathrm{Count}(w_k)$ is the number of agents activated for the dimension according to its weight. The final activation decision is the union of the two paths:

\begin{equation}
\mathrm{Activate}(P_i)=
\mathrm{Activate}_A(P_i)\lor\mathrm{Activate}_B(P_i)
\label{eq:activate_union}
\end{equation}

To control computational cost, the maximum number of activated agents is constrained by $K_{\max}=6$. If the union of the two activation paths exceeds this bound, the top $K_{\max}$ agents are retained according to $M(T,P_i)$. The activation procedure therefore uses four fixed hyperparameters: \(\theta_{\mathrm{score}}=100\), \(\theta_{\mathrm{dim}}=20\%\), the segmented rule below, and \(K_{\max}=6\):

\begin{equation}
\mathrm{Count}(w_k)=
\begin{cases}
0, & w_k<20\%,\\
1, & 20\%\leq w_k<30\%,\\
2, & 30\%\leq w_k<50\%,\\
3, & w_k\geq50\%
\end{cases}
\label{eq:count_rule}
\end{equation}

Threshold statistics, cross-review criteria, and fusion-strategy comparisons are provided in Supplementary Tables~S4--S7.

After activation, selected persona agents enter parallel generation, cross-review, disagreement convergence, and multi-level fusion. All activated agents receive the same task input, but their persona prompts induce different initial reasoning paths. Cross-review identifies invariant violations, causal gaps, parameter inconsistencies, and alternative proposals; unresolved disagreements trigger revision until convergence or the maximum number of rounds is reached.

The initial output of each activated persona agent is denoted as:

\begin{equation}
\begin{aligned}
S_i&=\operatorname{Agent}_i(T),
\{S_1,S_2,\ldots,S_n\}=\operatorname{ParallelProcess}\!\left(
T,\{\operatorname{Agent}_1,\ldots,\operatorname{Agent}_n\}
\right)
\end{aligned}
\label{eq:parallel_process}
\end{equation}

where $T$ is the input task, $\operatorname{Agent}_i$ is the $i$-th activated persona agent, $S_i$ is its initial solution, and $n$ is the number of activated agents.

The process runs for at most five rounds or stops earlier when the disagreement level satisfies the convergence condition. The agent cross-review criteria are provided in Supplementary Table S6, and the fusion-strategy comparison is provided in Supplementary Table S7.

The global disagreement level $D$ is computed as a weighted sum of the relative differences of current disputed variables:

\begin{equation}
D=\frac{1}{m}\sum_{j=1}^{m}w_j\delta_j
\label{eq:disagreement}
\end{equation}

Here, $m$ is the number of disputed variables, $w_j$ is the associated task-dimension weight, and $\delta_j$ is the maximum relative difference in the current round:

\begin{equation}
\delta_j=
\frac{\max_i(v_{i,j})-\min_i(v_{i,j})}
{\mathrm{range}_j}
\label{eq:relative_difference}
\end{equation}

where $v_{i,j}$ is the value assigned by agent $i$ to variable $j$, and $\mathrm{range}_j$ is its reasonable range. When $\delta_j>15\%$, variable $j$ is disputed. For high-risk tasks, routine engineering tasks, and conceptual exploration tasks, the convergence thresholds are $D<5\%$, $D<10\%$, and $D<15\%$, respectively. If convergence is not reached, unresolved variables remain as open issues with their rationales.

The multi-level fusion step first performs weighted fusion of quantitative or semi-quantitative information:

\begin{equation}
S_{\mathrm{weighted}}=\sum_{i=1}^{n}\tilde{w}_iS_i
\label{eq:weighted_fusion}
\end{equation}

where $\tilde{w}_i$ is the fusion weight and is distinguished from the task-dimension weight $w_i$. The fusion weight is determined by the agent's prior capability score, relevance to the current task, and match to the dominant dimension:

\begin{equation}
\tilde{w}_i=
\frac{
\mathrm{Score}_i\cdot\mathrm{Relevance}_i\cdot\mathrm{DimMatch}_i
}{
\sum_{j=1}^{n}
\mathrm{Score}_j\cdot\mathrm{Relevance}_j\cdot\mathrm{DimMatch}_j
}
\label{eq:fusion_weight}
\end{equation}

The second layer is dimension-expert adjudication: for qualitative conclusions or conflicting key judgments, $\mathrm{TopAgent}_k$ denotes the best-performing activated agent for task dimension $k$ and adjudicates the result. The third layer is consistency correction. If weighted fusion conflicts substantially with the dimension-expert judgment, the latter is prioritized and the conflict source is marked. Full formulas are provided in Supplementary Note~S3.

Finally, the system checks physical-law consistency, logical coherence, engineering feasibility, and safety compliance. Failed checks are marked as uncertainty or risk sources. Algorithm 1 gives the corresponding pseudocode.

\begin{algorithm}[htbp]
\caption{Dual-path on-demand activation and multi-level fusion}
\label{alg:dual_path_wakeup}
\begin{algorithmic}[1]
\Require Task $T$; task-weight vector $\mathbf{w}_{\mathrm{prob}}$; persona capability matrix $\mathbf{C}_{\mathrm{cap}}$; thresholds $\theta_{\mathrm{score}}$ and $\theta_{\mathrm{dim}}$; maximum activated agents $K_{\max}$
\Ensure Final fused answer $S_{\mathrm{final}}$
\State Map $\mathbf{w}_{\mathrm{prob}}$ to capability-demand vector $\mathbf{w}'$.
\For{each persona agent $P_i$}
\State Compute $M(T,P_i)=\sum_j w'_j c_{i,j}$.
\EndFor
\State Score-driven activation: $\mathcal{S}_A=\{P_i \mid M(T,P_i)\geq \theta_{\mathrm{score}}\}$.
\For{each task dimension $k$}
\If{$w_k\geq\theta_{\mathrm{dim}}$}
\State Select top $\mathrm{Count}(w_k)$ agents from $\mathrm{Pool}_k$.
\EndIf
\EndFor
\State Dimension-driven activation: $\mathcal{S}_B$ is the union of selected dimension agents.
\State Activated set: $\mathcal{A}_T=\mathcal{S}_A\cup\mathcal{S}_B$.
\If{$|\mathcal{A}_T|>K_{\max}$}
\State Retain the top $K_{\max}$ agents in $\mathcal{A}_T$ according to $M(T,P_i)$.
\EndIf
\State Parallel reasoning: $\{S_1,\ldots,S_n\}=\operatorname{ParallelProcess}(T,\mathcal{A}_T)$.
\State Agents review, revise, and iterate until $D$ satisfies the task-specific convergence threshold or the maximum number of rounds is reached.
\State Compute weighted fusion, apply dimension-expert adjudication, and perform consistency correction.
\State \Return $S_{\mathrm{final}}$ with uncertainty and unresolved-risk notes if needed.
\end{algorithmic}
\end{algorithm}

\subsection{Evaluation protocol}

All experiments use the same API-calling, parsing, scoring, and resource-statistics pipeline so that models, persona prompts, and collaboration structures are compared under consistent conditions. Supplementary Figure~S20 defines the standardized invocation and control framework, while Table~\ref{tab:model_basic_info} records the basic configuration of the five candidate Chinese large language models. Three benchmark resources serve different evaluation functions. The RFB-specific set evaluates specialized RFB reasoning and supports initial base-model selection. The ESS-LLM Benchmark provides the broader energy-storage task pool for persona-agent evaluation, capability-matrix calibration, and the ESS-LLM-derived generalization subset used in base-model selection. The 100-task core validation set compares fixed multi-agent configurations and on-demand activation, with 20 questions sampled from each of the five categories.

The primary scoring dimensions are professional knowledge accuracy, logical-analysis ability, and proposal innovation. Each dimension is scored on a 0-100 scale, and the overall score is computed as:

\begin{equation}
\mathrm{Score}
=0.30\times\mathrm{Knowledge}
+0.40\times\mathrm{Logic}
+0.30\times\mathrm{Innovation}
\label{eq:overall_score}
\end{equation}

Professional knowledge accuracy evaluates terminology, mechanisms, data, and domain facts; logical-analysis ability evaluates decomposition, causal chains, and multi-constraint consistency; and proposal innovation evaluates novelty and practical inspiration. Before scoring, problem invariants $\mathcal{I}$ are checked. Answers violating thermodynamic laws, electrochemical boundaries, safety red lines, or explicit engineering specifications are treated as factually incorrect and excluded from subsequent scoring across the three dimensions.

All model calls are made through a unified RESTful API. Responses are parsed into structured outputs for scoring, cross-review, and resource statistics, including problem analysis, key-constraint identification, proposal design, parameter suggestions, and feasibility assessment. The complete schema is provided in Supplementary Note~S4. Resource usage is recorded for each configuration, including token usage, inference time, memory usage, and average activated agents. These metrics compare output quality and collaboration cost for fixed and on-demand strategies.

\Needspace{0.28\textheight}
\begin{table}[!htbp]
\centering
\setcounter{table}{1}
\caption{\textbf{Basic information of the five candidate Chinese large language models.}}
\label{tab:model_basic_info}
\footnotesize
\setlength{\tabcolsep}{3pt}
\renewcommand{\arraystretch}{1.04}
\begin{adjustbox}{max width=0.96\textwidth,center}
\begin{tabular}{L{0.13\textwidth}*{5}{C{0.155\textwidth}}}
\toprule
\textbf{Item} & \textbf{DeepSeek} & \textbf{Qwen} & \textbf{GLM} & \textbf{Doubao} & \textbf{Kimi} \\
\midrule
\rowcolor{groupgray}
\multicolumn{6}{c}{\textbf{Foundation Model Information}}\\
Developer & DeepSeek & Alibaba Cloud & Zhipu AI & ByteDance & Moonshot AI \\
Version & DeepSeek-V3-Plus & Qwen2.5-Plus-72B & GLM-4-Plus & Doubao-Pro-2 & Kimi-K2 \\
Parameter Size & 686 B & 72 B & 176 B & 132 B & Undisclosed \\
Model Type & MoE + RL & Transformer & GLM & Transformer & MoE \\
\midrule
\rowcolor{groupgray}
\multicolumn{6}{c}{\textbf{Core Technical Parameters}}\\
Context Length & 128 K & 64 K & 128 K & 64 K & 256 K \\
Token Limit & 4 K & 8 K & 4 K & 8 K & 4 K \\
Response & \textless{} 1 s & \textless{} 1 s & \textless{} 1.5 s & \textless{} 0.8 s & \textless{} 1.2 s \\
\bottomrule
\end{tabular}
\end{adjustbox}
\end{table}

\FloatBarrier

\section{Results}

\subsection{Base-model selection and persona-prompting validation}

Base-model selection was first performed on the RFB-specific task set to ensure that subsequent persona-agent and multi-agent experiments were built on a sufficiently stable RFB reasoning capability. Five candidate LLMs--DeepSeek-V3-Plus, Qwen2.5-Plus-72B, GLM-4-Plus, Doubao-Pro-2, and Kimi-K2--were evaluated on the same 50 RFB-specific tasks under the same API settings, structured-output schema, and three-dimensional scoring protocol. The two-stage comparison in Table~\ref{tab:model_two_stage} identifies the strongest and most balanced candidate for the subsequent persona-agent experiments.

\begin{table}[!htbp]
\centering
\setcounter{table}{2}
\caption{\textbf{Comprehensive performance comparison of five candidate models on the two-stage test sets.}}
\label{tab:model_two_stage}
\small
\renewcommand{\arraystretch}{1.10}
\begin{adjustbox}{width=\textwidth,center}
\begin{tabular}{p{0.20\textwidth}ccccc}
\toprule
\textbf{Metric dimension} & \textbf{DeepSeek-V3-Plus} & \textbf{Qwen2.5-Plus-72B} & \textbf{GLM-4-Plus} & \textbf{Doubao-Pro-2} & \textbf{Kimi-K2} \\
\midrule
\rowcolor{groupgray}
\multicolumn{6}{c}{\textbf{Flow Battery Problem -- Average Scores by Evaluation Criteria}}\\
\makecell[l]{Knowledge Accuracy} & 91.2 & 87.6 & 87.1 & 84.3 & 85.3 \\
\makecell[l]{Logical Analysis} & 89.7 & 86.5 & 85.9 & 83.3 & 83.9 \\
\makecell[l]{Solution Innovation} & 87.3 & 85.3 & 86.5 & 83.4 & 83.6 \\
Average Score & 89.4 & 86.5 & 86.4 & 83.7 & 84.2 \\
\midrule
\rowcolor{groupgray}
\multicolumn{6}{c}{\textbf{ESS-LLM Benchmark -- Average Scores by Evaluation Criteria}}\\
\makecell[l]{Knowledge Accuracy} & 88.2 & 85.8 & 84.5 & 83.2 & 82.5 \\
\makecell[l]{Logical Analysis} & 87.5 & 85.2 & 84.0 & 82.5 & 81.8 \\
\makecell[l]{Solution Innovation} & 85.8 & 84.5 & 85.2 & 83.0 & 82.2 \\
Average Score & 87.2 & 85.2 & 84.5 & 82.9 & 82.2 \\
\midrule
\rowcolor{groupgray}
\multicolumn{6}{c}{\textbf{ESS-LLM Benchmark -- Average Scores by Problem Type}}\\
Materials R\&D & 88.5 & 86.2 & 85.0 & 83.5 & 82.8 \\
Stack Engineering & 86.8 & 85.5 & 84.2 & 83.0 & 82.5 \\
System Modeling & 87.2 & 85.8 & 86.5 & 83.5 & 83.0 \\
Safety Analysis & 86.5 & 84.8 & 84.0 & 84.5 & 83.2 \\
System Integration & 85.8 & 84.5 & 85.2 & 83.8 & 83.5 \\
\bottomrule
\end{tabular}
\end{adjustbox}
\end{table}

DeepSeek-V3-Plus achieved the highest overall score among the five candidate models and showed relatively balanced performance across professional knowledge, logical analysis, and proposal innovation. Its advantage was most relevant to this study in RFB tasks that required simultaneous consideration of electrochemical mechanisms, materials constraints, and system-level engineering boundaries. Therefore, DeepSeek-V3-Plus was selected as the shared base model for all subsequent persona-agent and multi-agent experiments.

The RFB-specific comparison included in Figure~\ref{fig:rfb_molecular_task} supports this choice by evaluating candidate models across task category, capability dimension, R\&D stage, and score stability. The comparison indicates that DeepSeek-V3-Plus was more balanced across the five RFB task categories and the three core capability dimensions and supports using DeepSeek-V3-Plus as the unified base model.

Beyond the RFB-specific evaluation, model selection was also checked using the broader ESS-LLM Benchmark categories for energy-storage consistency. The two-stage model comparison is summarized in Table~\ref{tab:model_two_stage}.

\begin{figure}[!htbp]
\centering
\includegraphics[width=0.86\linewidth,height=0.52\textheight,keepaspectratio]{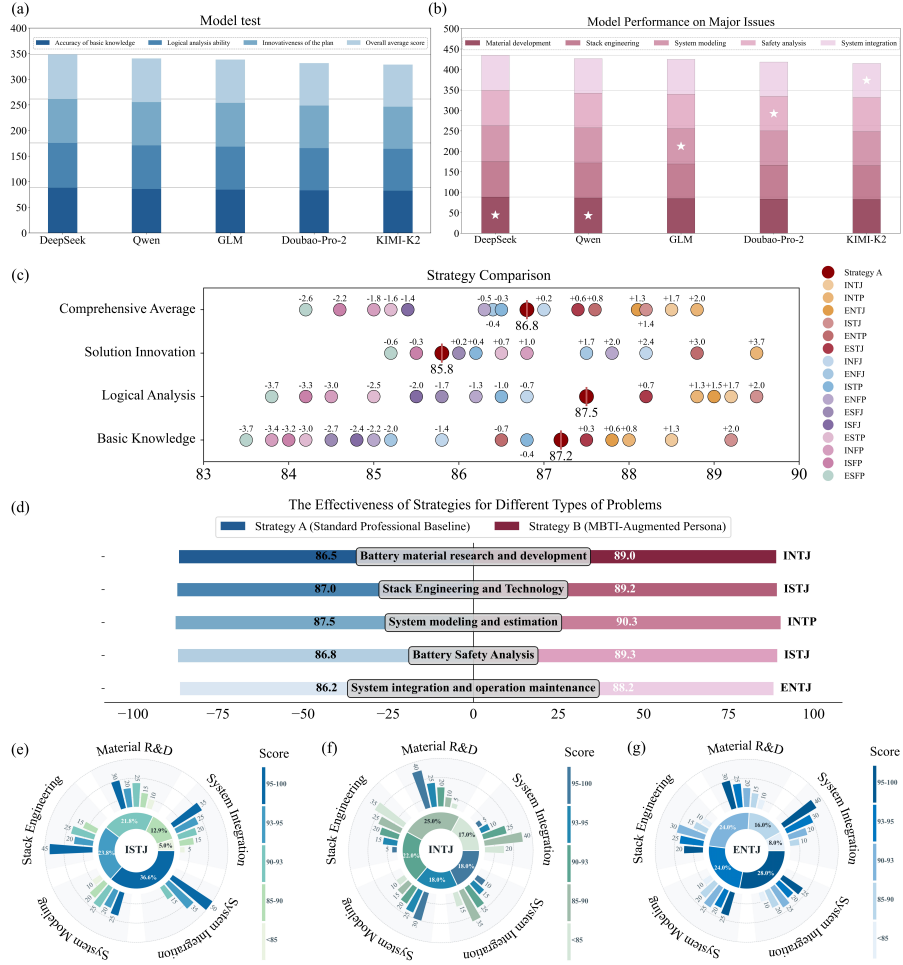}
\caption{\textbf{Model validation, persona effectiveness, and score distributions of the top-performing persona agents.} The merged figure summarizes model generalization, the task-dependent effect of MBTI-inspired persona prompting relative to the standard professional baseline, and score-distribution patterns for the top-performing persona agents.}
\label{fig:persona_validation}
\end{figure}

To evaluate whether the selected base model responds to persona prompts in a task-dependent and repeatable manner, this study compares Strategy A, the generic expert prompt, with Strategy B, the MBTI-augmented persona prompt. The experiment changes only the persona layer in the system prompt, while keeping the base model and scoring protocol unchanged. Figure~\ref{fig:persona_validation} combines model validation, persona effectiveness, and the score distributions of the top-performing persona agents.

Persona prompting affected not only wording but also reasoning organization and decision emphasis in some task categories. Some MBTI-inspired persona agents achieved higher overall scores than Strategy A. INTP and INTJ obtained overall improvements of 2.0 and 1.7 points, respectively. At the category level, the number of effective persona types was bounded: only three to five persona types produced gains above 0.5 points in each problem category. Materials R\&D and system-modeling tasks benefited more from intuitive-thinking persona prompts associated with abstract reasoning and mechanistic exploration, whereas safety-analysis tasks depended more on persona prompts emphasizing structured execution and risk-boundary identification. These findings support the task-dependent effect of MBTI-inspired persona prompting in energy-storage tasks, while also showing that persona augmentation should be treated as a conditional adaptation mechanism rather than a universally beneficial prompt strategy.

Based on these two evaluations, DeepSeek-V3-Plus was selected as the shared base model for all subsequent persona-agent and multi-agent collaboration experiments, and Strategy B was used to construct the 16 MBTI-inspired persona agents. The subsequent persona-level analysis examines their capability differentiation, cognitive advantage matrix, and single-persona coverage boundaries on the full ESS-LLM Benchmark.

\subsection{Persona-dependent cognitive advantages}

After selecting DeepSeek-V3-Plus and observing the task-dependent gains of Strategy B, this study evaluated 16 MBTI-inspired persona agents on the full 500-task ESS-LLM Benchmark. The purpose was not to validate MBTI as a psychometric instrument, but to evaluate whether different persona prompts induce distinguishable reasoning biases that adapt differently to energy-storage R\&D tasks. In other words, the 16 persona agents are treated as structured cognitive-bias templates that generate comparable reasoning preferences under the same model capability, task inputs, and scoring protocol.

\begin{table}[!tbp]
\centering
\setcounter{table}{3}
\caption{\textbf{Performance metrics of 16 MBTI-inspired persona agents across six capability dimensions.}}
\label{tab:persona_matrix}
\small
\renewcommand{\arraystretch}{1.08}
\begin{adjustbox}{width=\textwidth,center}
\begin{tabular}{cccccccc}
\toprule
\makecell{\textbf{MBTI}\\\textbf{Type}} & \makecell{\textbf{Innovation}\\\textbf{Insight}} & \makecell{\textbf{Material}\\\textbf{Design}} & \makecell{\textbf{Theoretical}\\\textbf{Modeling}} & \makecell{\textbf{Logical}\\\textbf{Completeness}} & \makecell{\textbf{Professional}\\\textbf{Accuracy}} & \makecell{\textbf{Safety}\\\textbf{Analysis}} & \makecell{\textbf{Overall}\\\textbf{Score}} \\
\midrule
ISTJ & 86.5 & 89.8 & 88.0 & 93.5 & 94.2 & 93.8 & 90.7 \\
INTJ & 91.8 & 92.5 & 89.5 & 90.2 & 89.8 & 88.5 & 90.4 \\
ENTJ & 89.2 & 91.2 & 90.0 & 90.5 & 89.8 & 89.5 & 90.4 \\
ESTJ & 85.8 & 90.8 & 88.2 & 92.5 & 93.0 & 92.5 & 89.9 \\
INTP & 92.5 & 90.8 & 94.2 & 90.5 & 88.5 & 87.8 & 89.4 \\
INFJ & 89.5 & 86.8 & 90.5 & 89.0 & 88.5 & 90.2 & 89.1 \\
ENTP & 93.0 & 89.5 & 92.5 & 88.5 & 87.0 & 86.8 & 88.5 \\
ISTP & 87.0 & 87.5 & 88.2 & 89.5 & 89.8 & 88.8 & 88.5 \\
ENFJ & 88.2 & 85.2 & 88.2 & 88.5 & 87.8 & 89.5 & 87.9 \\
ENFP & 91.0 & 85.5 & 89.8 & 87.0 & 86.2 & 87.8 & 87.9 \\
ESTP & 87.5 & 87.0 & 87.2 & 88.5 & 89.0 & 87.5 & 87.8 \\
ISFJ & 85.5 & 86.2 & 87.0 & 88.0 & 88.8 & 88.8 & 87.4 \\
ESFJ & 85.0 & 86.5 & 87.5 & 87.8 & 88.5 & 88.5 & 87.3 \\
INFP & 90.5 & 85.0 & 88.5 & 86.5 & 85.8 & 87.2 & 87.3 \\
ISFP & 88.5 & 85.0 & 86.8 & 86.8 & 86.5 & 87.0 & 86.8 \\
ESFP & 87.8 & 85.5 & 86.5 & 86.2 & 85.8 & 86.8 & 86.4 \\
\bottomrule
\end{tabular}
\end{adjustbox}
\end{table}

\begin{figure}[!htbp]
\centering
\includegraphics[width=0.96\linewidth,height=0.60\textheight,keepaspectratio]{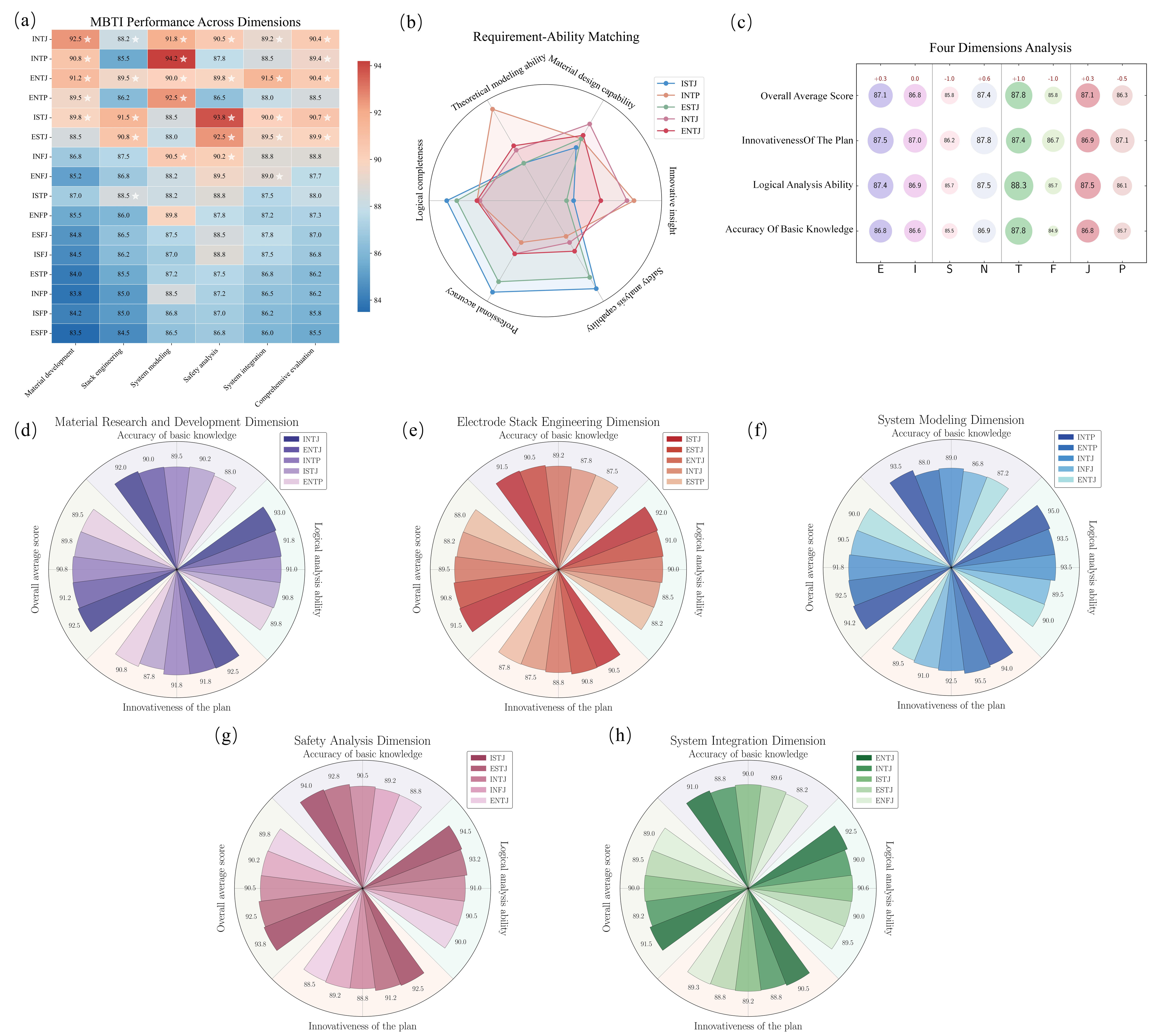}
\caption{\textbf{Cognitive-advantage matrix of 16 MBTI-inspired persona agents.} The figure shows task-category performance, capability-dimension differentiation, MBTI-letter aggregate trends, and top-persona performance patterns, demonstrating that persona prompts induce structured but task-dependent cognitive advantages. (a) Persona-task performance matrix across five energy-storage task categories; (b) top persona agents in six capability dimensions; (c) aggregated statistics based on MBTI single-letter dimensions, interpreted as prompt-induced cognitive-bias patterns rather than psychometric conclusions; (d-h) knowledge accuracy, logical reasoning, and proposal-innovation performance of top persona agents across task categories.}
\label{fig:persona_advantage}
\end{figure}

Table~\ref{tab:persona_matrix} and Figure~\ref{fig:persona_advantage}(a) jointly show that persona prompting did not produce a uniform gain across the 16 MBTI-inspired persona agents. Higher-performing agents were mainly found among the ISTJ, INTJ, ENTJ, ESTJ, and INTP types, but their advantages differed across materials R\&D, stack engineering, system modeling, safety analysis, and system integration. This pattern indicates that persona prompts mainly shifted reasoning priorities under different task demands rather than improving all categories uniformly.

The six-dimensional capability comparison in Figure~\ref{fig:persona_advantage}(b) further shows that no single persona agent dominated innovation insight, material-design ability, theoretical modeling, logical completeness, professional accuracy, and safety-analysis ability simultaneously. Agents associated with divergent exploration and abstract modeling performed better in innovation insight and theoretical modeling, whereas agents associated with structured execution and constraint checking were more stable in logical completeness, professional accuracy, and safety analysis. This differentiation supports the persona-capability matrix defined by the task-persona matching formulation and shows that the cognitive advantage matrix should be interpreted as a multidimensional capability structure rather than a single ranking.

The MBTI-letter aggregation in Figure~\ref{fig:persona_advantage}(c) provides a coarse view of prompt-induced cognitive bias. The T/F dimension suggests that logical filtering, factual consistency, and technical constraints tended to improve stability. The J/P dimension suggests that structured execution, boundary checking, and procedural closure may be useful for strongly constrained engineering tasks. The N/S dimension distinguishes abstract mechanistic exploration from concrete parameter checking. These results reflect output biases induced by prompt templates and do not constitute conclusions about real personality types.

The task-category panels in Figure~\ref{fig:persona_advantage}(d-h) explain why the top persona agents differ by domain. Materials-R\&D tasks involve molecular-structure hypotheses, performance-stability trade-offs, and multi-objective design, revealing advantages of abstract reasoning and exploratory persona agents. System-modeling tasks emphasize state variables, coupling relationships, and mechanistic abstraction, whereas stack engineering, safety analysis, and system integration contain more process constraints, failure boundaries, and engineering feasibility requirements. These patterns indicate that no single persona agent maintains a stable advantage across all task dimensions.

The score-interval distributions in Figure~\ref{fig:persona_validation} summarize the top-performing persona agents, while the complete task-specific blind-spot patterns of the top five single-persona agents across the 500 energy-storage tasks are provided in Supplementary Table~S8. Even the strongest overall persona agent retained low-score intervals and category-specific shortcomings. ISTJ still had 37\% of tasks below 93 points; INTJ had 64\% of tasks below 93 points, including 14\% below 85 points; and ENTJ and ESTJ each had no more than 60\% of tasks at or above 93 points. These distributions show that a high average score does not imply complete task coverage and that top persona agents still have single-persona blind spots.

Overall, the 16 MBTI-inspired persona agents exhibit a distinguishable cognitive advantage matrix in energy-storage R\&D tasks. Their advantages differ across task categories and capability dimensions, and no persona agent maintains an absolute advantage in all dimensions. Therefore, relying on the best single persona is insufficient for full cognitive coverage in multi-scale, multi-objective, and strongly constrained energy-storage tasks. Fixed multi-agent combination results are then used to evaluate whether complementary reasoning can compensate for these blind spots.

\subsection{On-demand dynamic activation}

\begin{figure}[!htbp]
\centering
\includegraphics[width=0.96\linewidth,height=0.62\textheight,keepaspectratio]{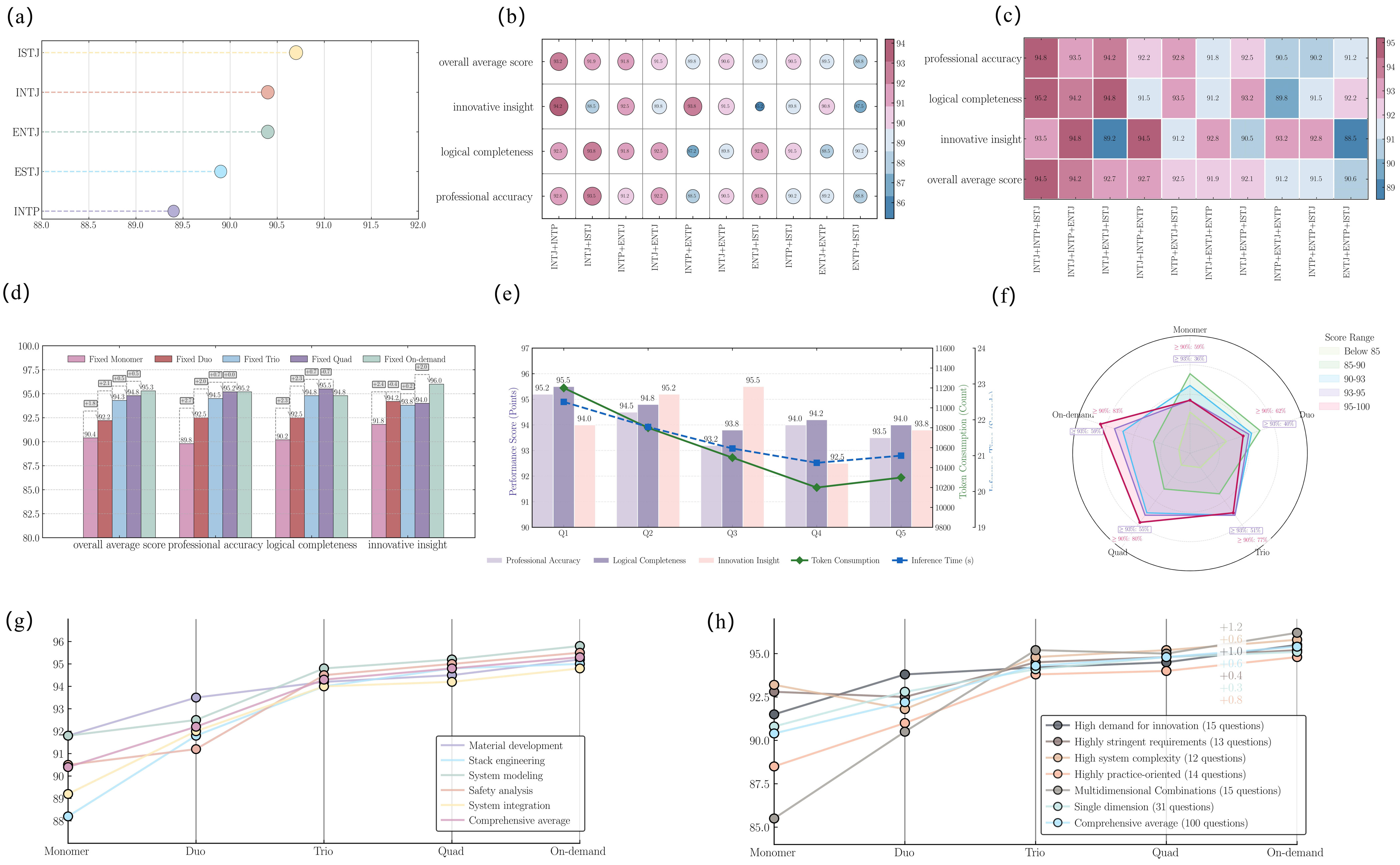}
\caption{\textbf{Fixed collaboration and on-demand dynamic activation performance.} The merged figure preserves the original fixed-combination and on-demand dynamic activation result panels (a)-(h): (a) top persona source pool, (b-d) fixed dyad, triad, and quartet collaboration results, (e) performance and resource evolution across fixed configuration scales, (f) score-interval distribution on the 100-task core validation set, and (g-h) performance benchmarks by energy-storage task type and cognitive-demand category. Complete fixed-combination and on-demand activation metrics are provided in Table~\ref{tab:fixed_dynamic_comparison} and Table~\ref{tab:activation_distribution}.}
\label{fig:fixed_dynamic_activation}
\end{figure}

After identifying persona-dependent cognitive advantages and single-persona blind spots on the ESS-LLM Benchmark, we first examined whether fixed persona combinations could provide complementary coverage. Figure~\ref{fig:fixed_dynamic_activation}(a-e) evaluates fixed dyad, triad, and quartet configurations selected from the high-performing persona pool and summarizes their performance/resource evolution. The INTJ + INTP dyad, INTJ + INTP + ISTJ triad, and INTJ + INTP + ISTJ + ENTJ quartet progressively improved task coverage, with overall scores increasing from 90.4 for the INTJ single-persona baseline to 92.2, 94.3, and 94.8, respectively (Table~\ref{tab:fixed_dynamic_comparison}).

Among fixed dyads, INTJ + INTP performed strongly, reflecting complementarity between structured abstraction and theoretical exploration. Adding ISTJ in the triad introduced stronger constraint checking and factual closure, while the quartet INTJ + INTP + ISTJ + ENTJ further added system-level coordination and proposal integration. However, the transition from triad to quartet mainly filled local gaps in an already covered capability structure rather than adding a new dominant capability dimension.

The bottleneck of fixed collaboration comes from its static configuration. All tasks call the same persona agents, which may introduce redundant review in narrow tasks and still suffer task-persona mismatch when task demand lies outside the fixed group's capability region. These results indicate that complementarity alone is insufficient; the participating persona agents must also be selected according to task demand.

To address this static-configuration bottleneck, ChargeBD applies on-demand dynamic activation on the 100-task core validation set. The mechanism dynamically selects MBTI-inspired persona agents according to task weights, persona matching scores, and key capability dimensions, followed by parallel generation, cross-review, and multi-level fusion.

Operationally, on-demand activation combines score-driven selection with dimension-driven compensation. Score-driven activation selects agents with high overall matching, while dimension-driven activation compensates for specialized demand in high-weight dimensions. Together, they allow a smaller collaboration scale for narrow tasks and more complementary agents for multi-dimensional tasks.

Table~\ref{tab:fixed_dynamic_comparison} provides the quantitative fixed-versus-dynamic comparison that supports this transition from static complementarity to adaptive coordination. The fixed quartet INTJ + INTP + ISTJ + ENTJ achieved an overall score of 94.8, whereas the On-Demand Dynamic configuration reached 95.3. It also reduced resource use from 11,200 to 5,800 tokens, from 22.5 to 11.2 s, and from 6.2 to 2.9 GB, with an average of 2.6 activated agents. Relative to the fixed quartet, token usage decreased by 48.2\%, inference time decreased by 50.2\%, and resource efficiency increased from 40\% to 82\%.

\begin{table}[!htbp]
\centering
\setcounter{table}{4}
\caption{\textbf{Comprehensive performance comparison between fixed and on-demand dynamic configurations.}}
\label{tab:fixed_dynamic_comparison}
\small
\renewcommand{\arraystretch}{1.10}
\begin{adjustbox}{width=\textwidth,center}
\begin{tabular}{lccccc}
\toprule
\multirow{2}{*}{\textbf{Configuration}} & \multicolumn{5}{c}{\textbf{Performance Metrics}}\\
\cmidrule(lr){2-6}
& \makecell{\textbf{Overall}\\\textbf{Score}} & \makecell{\textbf{Token}\\\textbf{Consumption}} & \makecell{\textbf{Inference}\\\textbf{Time (s)}} & \makecell{\textbf{Memory}\\\textbf{Usage (GB)}} & \makecell{\textbf{Average Active}\\\textbf{Agents}}\\
\midrule
INTJ (Single) & 90.4 & 4,200 & 8.2 & 2.1 & 1.0 \\
INTJ+INTP (Pair) & 92.2 & 6,300 & 12.5 & 3.2 & 2.0 \\
INTJ+INTP+ISTJ (Trio) & 94.3 & 8,400 & 16.8 & 4.5 & 3.0 \\
INTJ+INTP+ISTJ+ENTJ (Quartet) & 94.8 & 11,200 & 22.5 & 6.2 & 4.0 \\
\rowcolor{rowgray}
On-Demand Dynamic & 95.3 & 5,800 & 11.2 & 2.9 & \textbf{2.6} \\
\midrule
\multirow{2}{*}{\textbf{Configuration}} & \multicolumn{5}{c}{\textbf{Efficiency and Score Distribution}}\\
\cmidrule(lr){2-6}
& \makecell{\textbf{Resource}\\\textbf{Efficiency}} & \makecell{\textbf{95--100}\\\textbf{Score Ratio}} & \makecell{\textbf{93--95}\\\textbf{Score Ratio}} & \makecell{\textbf{90--93}\\\textbf{Score Ratio}} & \makecell{\textbf{85--90}\\\textbf{Score Ratio}}\\
\midrule
INTJ (Single) & 45\% & 18\% & 36\% & 59\% & 24\% \\
INTJ+INTP (Pair) & 58\% & 19\% & 40\% & 62\% & 25\% \\
INTJ+INTP+ISTJ (Trio) & 49\% & 25\% & 51\% & 77\% & 18\% \\
INTJ+INTP+ISTJ+ENTJ (Quartet) & 40\% & 29\% & 55\% & 80\% & 15\% \\
\rowcolor{rowgray}
On-Demand Dynamic & 82\% & 32\% & 59\% & 83\% & 13\% \\
\bottomrule
\end{tabular}
\end{adjustbox}
\caption*{\footnotesize Note: The score-threshold columns are cumulative proportions except the 85--90 column, which denotes the proportion of samples in the 85--90 score interval. The average number of activated agents for On-Demand Dynamic is task-dependent and equals 2.6 on average.}
\end{table}

\begin{table}[!htbp]
\centering
\setcounter{table}{5}
\caption{\textbf{Activation-configuration distribution of on-demand dynamic activation combinations.}}
\label{tab:activation_distribution}
\small
\renewcommand{\arraystretch}{1.10}
\begin{adjustbox}{width=\textwidth,center}
\begin{tabular}{lccccc}
\toprule
\textbf{Activation Configuration} & \makecell{\textbf{Active Agents}} & \makecell{\textbf{Frequency}} & \makecell{\textbf{Proportion}} & \makecell{\textbf{Mean}\\\textbf{Score}} & \makecell{\textbf{Token Change}\\\textbf{vs. Quartet}} \\
\midrule
INTJ & 1 & 1 & 1\% & 93.5 & -9,000 \\
INTP & 1 & 1 & 1\% & 93.0 & -9,000 \\
INTJ+INTP & 2 & 20 & 20\% & 95.0 & -7,000 \\
INTJ+ISTJ & 2 & 12 & 12\% & 94.5 & -7,000 \\
INTP+ENTJ & 2 & 8 & 8\% & 95.0 & -7,000 \\
ISTJ+ESTJ & 2 & 8 & 8\% & 94.8 & -7,000 \\
INTJ+INTP+ISTJ & 3 & 18 & 18\% & 95.8 & -5,000 \\
INTJ+INTP+ENTJ & 3 & 12 & 12\% & 95.5 & -5,000 \\
INTJ+ISTJ+ENTJ & 3 & 8 & 8\% & 95.5 & -5,000 \\
INTJ+INTP+ISTJ+ENTJ & 4 & 12 & 12\% & 96.2 & 0 \\
\rowcolor{rowgray}
\textbf{Total} & \textbf{2.6} & \textbf{100} & \textbf{100\%} & \textbf{95.3} & \textbf{-5,400} \\
\bottomrule
\end{tabular}
\end{adjustbox}
\caption*{\footnotesize Note: All configurations are dynamically activated based on task requirements. $\Delta$ Tokens represents the change in token consumption relative to the fixed four-agent combination (INTJ+INTP+ISTJ+ENTJ). The number of activated agents in the total row (2.6) represents the weighted average of all 100 test samples.}
\end{table}

Table~\ref{tab:activation_distribution} reports the activation-configuration distribution, showing that the advantage of on-demand activation does not come from simply increasing the number of agents, but from finer task-capability matching. The fixed quartet calls four persona agents for every task, whereas the on-demand configuration adjusts the activation scale, excludes low-relevance agents, and reduces redundant review and integration cost. For multi-dimensional tasks, it can still activate multiple complementary agents to maintain cognitive coverage.

Figure~\ref{fig:fixed_dynamic_activation}(f) supports the score-distribution interpretation on the 100-task core validation set. Compared with the fixed quartet, on-demand activation placed more tasks in the 95-100 interval (32 vs. 29), increased the proportion at or above 93 points (59\% vs. 55\%), and reduced tasks below 85 points (4 vs. 5). This suggests improved average score, broader high-quality output coverage, and lower tail risk.

Panels (g,h) in Figure~\ref{fig:fixed_dynamic_activation} further support the adaptation argument by grouping the results by task type and cognitive demand. For high-innovation-demand tasks, the system tended to activate agents associated with abstract exploration, mechanistic reasoning, and theoretical modeling. For high-constraint or practice-oriented tasks, it tended to include agents emphasizing logical completeness, professional accuracy, and risk-boundary identification. Compared with fixed combinations, this task-level selection mechanism aligned prompt-induced cognitive biases more closely with current problem demand.

Overall, on-demand dynamic activation improved the performance-resource trade-off of fixed collaboration. It maintained and slightly exceeded the fixed quartet score while reducing token usage, inference time, memory usage, and average participating agents. These findings suggest that MBTI-inspired persona agents are useful not only as diverse cognitive templates, but also as templates that can be organized through task-persona matching. The ChargeBD Web Platform is then used to demonstrate how ChargeBD can be exposed as a transparent and auditable reasoning process.

\subsection{ChargeBD Web Platform workflow demonstration}

The benchmark analyses above provide offline evidence for base-model selection, persona-agent differentiation, fixed-collaboration behavior, and on-demand dynamic activation. The ChargeBD Web Platform, evolved from the earlier MBTI-Agent-Cluster prototype, implements the proposed reasoning workflow as an interactive and auditable interface. The system complements benchmark evaluation by exposing process-level behavior, including task decomposition, persona selection, persona-specific rationale summaries, and final synthesis.

The workflow system uses a React 19, TypeScript, and Vite frontend, a Node.js and Express backend, and a DeepSeek-compatible API layer. The interface contains a conversation-history sidebar, chat workspace, workflow timeline, and expert-library overlay showing 16 MBTI-inspired persona agents, allowing users to inspect the interaction state and persona-agent pool rather than only a final answer.

The implemented workflow follows five stages. First, the user problem is analyzed through five-dimensional task-weight decomposition covering materials R\&D, stack engineering, system modeling, safety analysis, and system integration. Second, the system performs dual-path on-demand activation using a matching-score path and a dimension-driven activation path. Third, selected persona agents generate concurrent persona-specific rationale and answer streams. Fourth, a chief-engineer synthesis stage integrates the selected expert outputs into a unified answer. Fifth, the system supports multi-turn interaction by retaining conversation history and carrying forward concise context from earlier turns.

The workflow system also makes activation inspectable. The dispatch card presents five-dimensional weights, Path A and Path B activation routes, selected-agent rationales, persona roles, matching scores, and trigger conditions. During generation, the interface separates intermediate rationale summaries from final answer content.

Two interaction modes are implemented. In collaborative mode, the system performs on-demand persona activation and multi-agent synthesis, corresponding to the fixed-collaboration and dynamic-activation analyses above. In single-expert mode, the user can specify one MBTI persona agent, providing an interface-level analogue of the single-persona comparisons. The benchmark scores reported above are produced by the offline evaluation protocol, not by the interactive workflow system.

The ChargeBD Web Platform should therefore be interpreted as an operational workflow demonstrator and inspection interface for ChargeBD rather than as an experimental battery-validation platform.

\section{Discussion}

This study evaluates ChargeBD as a persona-conditioned heterogeneous-agent reasoning framework for constrained energy-storage R\&D tasks. Starting from RFB research, the work builds an RFB-specific task set and expands it into the ESS-LLM Benchmark to test reasoning across molecular design, materials, components, operation, system management, and safety. Within this setting, the results support three connected observations: DeepSeek-V3-Plus provides a stable shared base model, persona prompting produces task-dependent reasoning patterns, and dynamic activation can improve the trade-off between answer quality and computational resources.

The persona component should be interpreted narrowly. MBTI-inspired persona agents are prompt-defined reasoning templates, not psychometric instruments and not representations of real human or model personality. Their role is to induce controlled variation in reasoning preference, risk orientation, task priority, and output organization while holding the base model, task input, and evaluation protocol constant. This framing allows persona effects to be compared as prompt-conditioned reasoning patterns rather than psychological attributes.

The persona-dependent results indicate that no single reasoning template is uniformly optimal. The 16 persona agents form task-dependent cognitive-advantage patterns across task categories and capability dimensions. This heterogeneity is important for RFB and broader energy-storage tasks because correct answers often require simultaneous attention to materials exploration, mechanistic plausibility, engineering constraints, safety limits, and system-level trade-offs. A high average score can therefore mask local blind spots, especially when a task fails through a constraint violation rather than a lack of general knowledge.

The benchmark design helps expose this issue because the tasks are not framed as generic question answering. They require the system to balance domain knowledge with engineering constraints and to produce answers that remain reviewable under a structured scoring protocol. The RFB-specific task set anchors the evaluation in a concrete battery domain, while the broader ESS-LLM Benchmark tests whether the same reasoning workflow can generalize across related energy-storage categories. This design does not remove evaluator dependence, but it provides a consistent setting for comparing single-persona, fixed-group, and dynamically activated reasoning.

Fixed multi-agent groups partially address this limitation by combining complementary reasoning patterns. Dyads, triads, and quartets improve coverage relative to single-persona baselines, but their marginal gains diminish as group size increases. A fixed quartet also uses the same participants for every task, even when the task emphasizes a narrow capability or requires a different balance of exploration, verification, and system reasoning. This explains why larger fixed groups can add redundancy without fully resolving task-persona mismatch.

On-demand dynamic activation is therefore the main structural contribution of ChargeBD. The mechanism combines task-persona matching with score-driven and dimension-driven activation, then uses parallel generation, cross-review, disagreement convergence, and multi-level fusion to synthesize selected outputs. The score-driven path favors agents with strong overall task fit, whereas the dimension-driven path retains agents aligned with high-weight task demands. This dual-path design is useful for energy-storage reasoning because a task may depend on a specialized constraint, such as safety, system integration, or mechanistic consistency, even when the best global matching score points elsewhere.

On the 100-task core validation set, dynamic activation achieved a higher overall score than the fixed quartet while reducing token use by 48.2\% and inference time by 50.2\%, with 2.6 activated agents on average. These results suggest that multi-agent value depends on selective coordination, not merely on increasing the number of agents. The result also clarifies why resource-aware activation matters for scientific reasoning workflows. If additional agents are called without task alignment, the system may increase cost and latency while adding overlapping arguments. By contrast, task-adaptive activation aims to preserve complementary reasoning while avoiding unnecessary generation.

The ChargeBD Web Platform provides an implementation-level view of this workflow. It exposes task decomposition, persona activation, concurrent persona reasoning, and chief-engineer synthesis as a transparent and auditable workflow system. This interface does not generate the benchmark scores and should not be read as a battery-performance validation platform. Its function is to make the reasoning process inspectable, so that candidate hypotheses, constraints, risks, and synthesized recommendations can be reviewed by domain experts.

This process-level visibility is part of the contribution, but it should be separated from evidence of battery performance. The Web Platform shows how task decomposition, persona selection, parallel reasoning, and synthesis can be exposed to users in a reviewable interface. It does not show that a proposed molecule, electrolyte formulation, electrode structure, operating strategy, or system-control recommendation will perform well in practice. Those claims require expert review, simulation, or experimental assessment outside the present computational benchmark.

Several boundaries remain. The benchmark and Web Platform provide computational reasoning-support evidence, not experimental validation of battery performance. Task annotation, task-weight assignment, scoring dimensions, and structured-output requirements may influence the observed patterns. Because all persona agents use the same shared base model, cross-model replication is needed before treating the persona-task matching patterns as model-independent. The current Web Platform also does not include a closed electrochemical simulation loop, an experimental feedback loop, or a separate multi-round agent-to-agent review protocol.

Overall, ChargeBD offers a resource-aware workflow for organizing heterogeneous LLM reasoning under energy-storage constraints. Its present evidence supports task-adaptive persona selection, transparent collaboration, and auditable synthesis, but downstream battery-development decisions require further validation. Future work should release and stress-test the benchmark, prompt templates, and scoring protocol; evaluate the framework across base models and engineering-science domains; and connect ChargeBD outputs with expert review, electrochemical simulation, or experimental assessment before any measurable R\&D claim is made.

\end{document}